\documentclass[conference]{IEEEtran}
\usepackage[pdftex]{graphicx}
\usepackage{paralist}
\usepackage{multirow}
\usepackage{tabularx}
\usepackage{url}

\usepackage{acronym}

\acrodef{sos}[SoS]{System of Systems}
\acrodefplural{sos}{Systems of Systems}
\acrodef{cs}[CS]{constituent system}
\acrodef{tl}[TL]{Transport Layer}
\acrodef{le}[LE]{Leader Election}

\def\contractspattern{Contract Pattern}
\def\fmconpatt{extended Contract Pattern}
\def\bo{B\&O}

\begin{document}

\title{Fault Modelling in System-of-Systems Contracts}

\author{\IEEEauthorblockN{Zoe Andrews, Jeremy Bryans, Richard Payne}
\IEEEauthorblockA{School of Computing Science\\
Newcastle University, UK\\
\{\emph{firstname.lastname}\}@ncl.ac.uk}
\and
\IEEEauthorblockN{Klaus Kristensen}
\IEEEauthorblockA{Bang \& Olufsen, Denmark\\
krt@bang-olufsen.dk}
}

\maketitle

\begin{abstract}
The nature of Systems of Systems (SoSs), large complex systems composed of independent, geographically distributed and continuously evolving constituent systems, means that faults are unavoidable.  Previous work on defining contractual specifications of the constituent systems within SoSs does not provide any explicit consideration for faults.  In this paper we address that gap by extending an existing pattern for modelling contracts with fault modelling concepts.  The proposed extensions are introduced with respect to an Audio Visual SoS case study from Bang \& Olufsen, before discussing how they relate to previous work on modelling faults in SoSs.

\medskip

\end{abstract}

%
\IEEEpeerreviewmaketitle


\section{Introduction}
\label{sec:intoduction}



\acp{sos} are an active area of research due to the complexity that is present as a consequence of several characteristics common to \acp{sos}~\cite{Maier98}; including: the independence of the \acp{cs} making up an \ac{sos}, the distribution of \acp{cs}, the evolution and reconfiguration of an \ac{sos}, and the emergent behaviours of an \ac{sos}. In addition to these characteristics (and conceivably due to them), Kopetz states that faults -- considered exceptional in traditional monolithic systems -- are normal and to be expected in \acp{sos}~\cite{Kopetz2013}. 
  We therefore need methods to assist in understanding and designing for faults in \acp{sos}.%

%

The job of the \ac{sos} integrator is to integrate ``disparate independent [constituent] systems up to an interoperable formation of systems'' and it is often a ``complex, risky, long, and frustrating effort''~\cite{Mordecai&13}. 
This is  due in large part to ``emergence'': a natural feature of \ac{sos}~\cite{Baldwin&13}.  We define \emph{emergence} in accordance with Nielsen et al. as ``the behaviours that arise as a result of the synergistic collaboration of constituent [systems]''~\cite{Nielsen2013}.  Emergence may be desired or undesired: the  \ac{sos} integration must  produce the desired emergent behaviour while preventing undesired emergent behaviour within the  \ac{sos}. 
Commonly \acp{cs} are not under the control of the \ac{sos} integrator, and thus we propose a contractual description of these \acp{cs} that captures their external behaviour~\cite{Bryans2014, Bryans2014a}. 

In this work, a contractual description (or \emph{contract}) defines the behaviour of a \ac{cs} in terms of its functionality and ordering of events. 
Currently, however, contracts contain no support for explicitly recording and dealing with faults.
Here we extend the  definition of contracts to enable the explicit recording of faults and fault tolerance in \acp{sos}.
We do this by building on two existing pieces of work: on contracts and on fault modelling. 

We envisage these extensions being used by two groups: \ac{cs} designers and \ac{sos} integrators.
The designer of a \ac{cs} (or its contract) may identify and acknowledge any deviations from desired behaviour of \acp{cs}. 
They may also state any such deviations from their environment (for example from other \acp{cs} in the \ac{sos}) that they mitigate. 
These extended contract definitions may be used by the \ac{sos} integrator as inputs to the integration process.  The \ac{sos} integrator may also observe deviations of the \acp{cs} not under their control, and wish to mitigate them. 
 Fault modelling techniques can be applied that explicitly state what deviations may be present in the \acp{cs} of the \ac{sos}, and identify those \acp{cs} that may be affected by or mitigate those deviations.  

In this paper, we propose extensions to an existing method for defining contracts; providing traceable fault modelling techniques to understand the effects of faults in contracts.  This allows: 
\begin{inparaenum}
\item[(i)] \ac{cs} designers to annotate their \ac{cs} contracts with failure modes and fault masking capabilities; 
\item[(ii)] \ac{sos} integrators to describe faults identified in the \acp{cs}; and
\item[(iii)] \ac{sos} integrators to define the means to recover from, or mitigate, the effects of \ac{sos} faults.
\end{inparaenum} 

In Section~\ref{sec:background} we go into more detail on relevant  background on modelling contracts and faults before describing related work in Section~\ref{sec:related}.  
In Section~\ref{sec:case-study} we propose extensions to the \contractspattern\ to include faults.  
These extensions are elicited using an Audio Visual \ac{sos} case study.  
We discuss how the extended contract pattern relates to our previous work on fault modelling in Section~\ref{sec:fm-contracts-fmaf}.  
Further work and our conclusions are presented in Section~\ref{sec:conclusions}.


\section{Background}
\label{sec:background}

Our work builds on the established field of design patterns.  Such patterns were originally conceived for designing buildings~\cite{Alexander&77}; their usage has since expanded to software design patterns~\cite{Gamma&95} (particularly for object-oriented software engineering).  Recent research provides a number of modelling patterns for \ac{sos} architectures~\cite{COMPASSD22.3}.  These were described using a structured approach that is underpinned by an ``architectural framework'' known as the CAFF~\cite{COMPASSD21.2}.  The CAFF provides a consistent way of describing both \emph{enabling patterns} (``patterns whose use enables a number of systems engineering applications''~\cite{COMPASSD22.3}) and \emph{architectural frameworks} (which ``provide guidance on information sets, or views, that may be used to present architectural information in a standard way''~\cite{COMPASSD22.1}).

This paper builds on two main areas of background work --  the \contractspattern~\cite{Bryans2014a} and the Fault Modelling Architectural Framework (FMAF)~\cite{Andrews2014}, both of which have been developed using the CAFF.  These are described in more depth in Sections~\ref{sec:contracts} and~\ref{sec:fmaf} respectively.

Both the \contractspattern\ and the FMAF are defined and demonstrated using SysML~\cite{OMG12SysML}. They are, however, notation agnostic and several notations have been developed for defining system architectures\footnote{In~\cite{Bryans&13} we survey this area more comprehensively, with particular emphasis on the support for the rigorous definition of interface specification.}. We use SysML in our work and in this paper primarily due to its increased use in industry.  The profiling features of SysML allow us to extend the base notation with the required concepts for the relevant domain.

%
%

\subsection{Contracts for \ac{sos}}
\label{sec:contracts}


We have previously defined the \contractspattern; a modelling pattern for defining contracts in \acp{sos}~\cite{Bryans2014a}. The \contractspattern\ is considered an enabling pattern in relation to our previous work on modelling patterns~\cite{COMPASSD22.3} -- it may be applied in many different domains and across different architectural patterns, for example service-oriented architecture, centralised etc.

The \contractspattern\ enables the analysis of emergent behaviours of an \ac{sos}, through the modelling of the internal behaviours of \acp{cs}. The pattern allows a modeller to define this behaviour in such a way as not to over-constrain the \ac{cs} implementations. This limited internal behaviour is considered a contract.  A \ac{cs} implementation is expected to conform to one or more contracts.

The \contractspattern\ comprises several related viewpoints, defining:
\begin{inparaenum}
\item[(i)] the contracts of the \ac{sos}; 
\item[(ii)] the conformance of the \acp{cs} to the contracts; 
\item[(iii)] the connections between the ports governed by contracts; 
\item[(iv)] the contract definitions; and
\item[(v)] the protocols constraining the behaviour of the contract. 
\end{inparaenum}
The viewpoints of the \contractspattern\ 
are identified in Table~\ref{tab:pattern_changes}
and are described in more detail in~\cite{Bryans2014a}.  They assume the use of the Interface Pattern, as defined in~\cite{COMPASSD22.3}, when considering the connections between the ports and interfaces governed by contracts.

We propose that contracts defined using the \contractspattern\ may be translated to the Compass Modelling Language (CML): a formal modelling language designed for SoSs~\cite{Bryans&13}.  Conformance relationships identified between CSs and contracts may be checked in terms of CML trace refinement.  Some technical challenges relating to this are discussed in~\cite{Bryans2014a}.

\subsection{Fault Modelling in \ac{sos}}
\label{sec:fmaf}


In our previous work we developed a Fault Modelling Architectural Framework (FMAF) for designing fault-tolerant SoSs \cite{Andrews2013, Andrews2014, Andrews2013c} and demonstrated how fault-tolerant properties of such models could be verified \cite{Andrews2013a}\footnote{The FMAF is considered an architectural framework, rather than a pattern, as it is of a larger scale than a pattern, and is designed to cover the full functionality of an SoS.}.  The FMAF defines a set of viewpoints that prompts an \ac{sos} developer to consider the impact of faults at the early stages of design, resulting in a coherent set of views that aid the stakeholders of the \ac{sos} to understand its erroneous and recovery behaviour.

The FMAF has been developed with respect to established dependability concepts \cite{Avizienis2004}.  An \ac{sos} \emph{failure} \cite{Andrews2014} is defined as a deviation of the service provided by the \ac{sos} from expected (correct) behaviour. An \emph{error} is defined as the part of the \ac{sos} state that can lead to its subsequent service failure. The adjudged or hypothesised cause of an error is called a \emph{fault} and (in keeping with our chosen nomenclature) a failure of a \ac{cs} can cause a fault of the \ac{sos}.

The set of viewpoints prescribed by the FMAF provide an approach for defining: 
\begin{inparaenum}
\item[(i)] faults, errors and failures of \acp{sos};
\item[(ii)] relationships between faults, errors and failures and \acp{cs}; 
\item[(iii)] structural designs that enable fault tolerance;
\item[(iv)] the behaviour of the \ac{sos} in the presence of errors; and
\item[(v)] the recovery behaviour provided by \acp{cs}.
\end{inparaenum}
For further details of these viewpoints, see \cite{Andrews2013c}.


\section{Related Work}
\label{sec:related}


A common theme of both the work on contracts and the work on fault modelling is the value of precision in the definition of interfaces. 
The SysML language allows basic operation signatures to be defined at interfaces, and pre- and postconditions, though these are rarely used in practice. 
The Design by Contract (DbC) software engineering technique~\cite{Meyer88} is built upon to constrain operations in the contract and interface definitions. DbC has been used in several other areas including software engineering design patterns~\cite{Champlain97} and in service-based components~\cite{Beugnard&99}. 
Another approach to contract design has been proposed in~\cite{Arnold&13}.  There contracts specify goals and requirements of an SoS, and are given an LTL semantics.
Previous work has considered: nonfunctional properties and DbC in architectural interfaces~\cite{Payne&10}; how interfaces in SysML can be translated into a formal notation (CML)~\cite{Bryans&13}; and a method of extending SysML interface descriptions with a contractual pattern~\cite{Bryans2014}.  The current research builds on this previous work.

Kopetz identifies a variety of potential issues and design principles for the control of cognitive complexity of \acp{sos}~\cite{Kopetz2013}.
Relied Upon Message Interfaces (RUMIs), similar to the notion of architectural interfaces, are used to specify the exchange of messages between \acp{cs} in terms of their ``syntax, semantics and temporal behaviour''~\cite{Kopetz2013}. Our approach aims to avoid the situation where there are ``multiple different semantic specifications of the same RUMI''~\cite{Kopetz2013}  (one for each \ac{cs} involved) by providing a standard approach for defining the behaviour of \acp{cs} through contract specifications.  By providing a notation to define faults and recovery within the contracts pattern we provide support for the identified need to consider ``fault containment and error propagation when defining RUMIs''~\cite{Kopetz2013}.

These are many examples of architecting fault-tolerant systems, however to the best of our knowledge this is limited to the systems level and not in the field of \ac{sos} engineering.  Relevant work includes: the Error Model Annex for AADL that allows architectural modelling of dependability features~\cite{Rugina&08}; the use of UML for modelling erroneous behaviour in embedded systems~\cite{Bernardi&07, Bondavalli&01}; and modelling using SysML for dependable complex physical systems~\cite{David&10} and to verify the safety requirements in the design of embedded safety-critical control systems~\cite{Petin&10}.


\section{Modelling Faults in Contract Descriptions}
%
%
\label{sec:case-study}


Bang \& Olufsen (B\&O) develop home Audio Visual (AV) networks in which several devices cooperate to provide a seamless experience for the user.   
The AV network exhibits many  properties typical of an SoS~\cite{Maier98, Nielsen2013}; for example, the devices (\acp{cs}) are \emph{heterogeneous}, with (potentially) a wide variation in \emph{autonomy}.  The \acp{cs} may include non-\bo\ products not fully under the control of the SoS. They may be capable of operating independently in a stand-alone mode. Although not distributed widely, the fact that \acp{cs} are not co-located offers the possibility of certain \emph{emergent} behaviours, such as sound which ``follows'' the listener round the home.  \emph{Evolution} may be present in the AV SoS as a result of  software or firmware upgrades.  

\begin{figure}[h]
	\centering
	\includegraphics[width=0.49\textwidth]{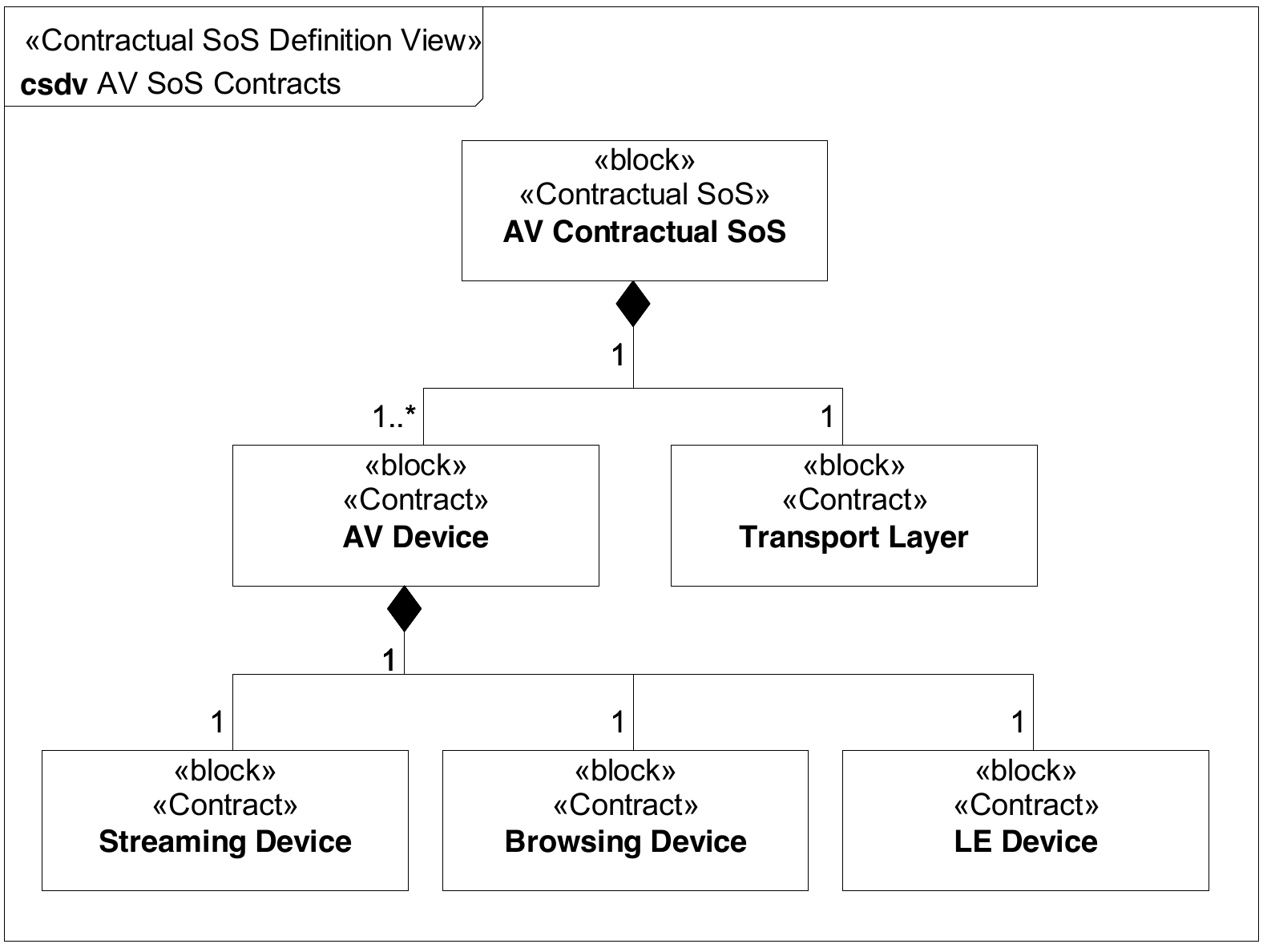}
	\caption{Contractual SoS Definition View of the AV SoS}
	\label{fig:avsosnominal}
\end{figure}

In order for the CSs to cooperate effectively they need to establish a single leader to coordinate the SoS.  The leader is established through an election process, which involves a series of messages being exchanged between the devices according to a predefined algorithm.  For the purposes of this illustrative example we focus on the election functionality of the SoS. To this end, for each \emph{AV Device} we define several contracts including the Leader Election Device (\emph{LE Device}) contract, which specifies the behaviour of the election functionality of an \emph{AV Device}.  The composition of contracts making up the SoS is shown in Figure~\ref{fig:avsosnominal}.

The SoS Leader Election protocol~\cite{Bryans2014,Bryans2014a} also requires a CS for transmitting messages -- the behaviour of such a CS is defined in the \emph{\ac{tl}} contract.   The reliable transmission of messages is a key aspect of the election.  For example, if messages to and from one CS are dropped then 
that CS may choose to elect a different leader than the rest of the SoS, resulting in an inconsistent AV experience.  As the network responsible for transmitting messages is an independent part of the SoS, only limited guarantees of its reliability can be made.  Therefore it is important to build fault tolerance into the transmission of messages to tolerate \emph{\ac{tl}} failures.

\begin{figure}[h]
	\centering
	\includegraphics[width=0.49\textwidth]{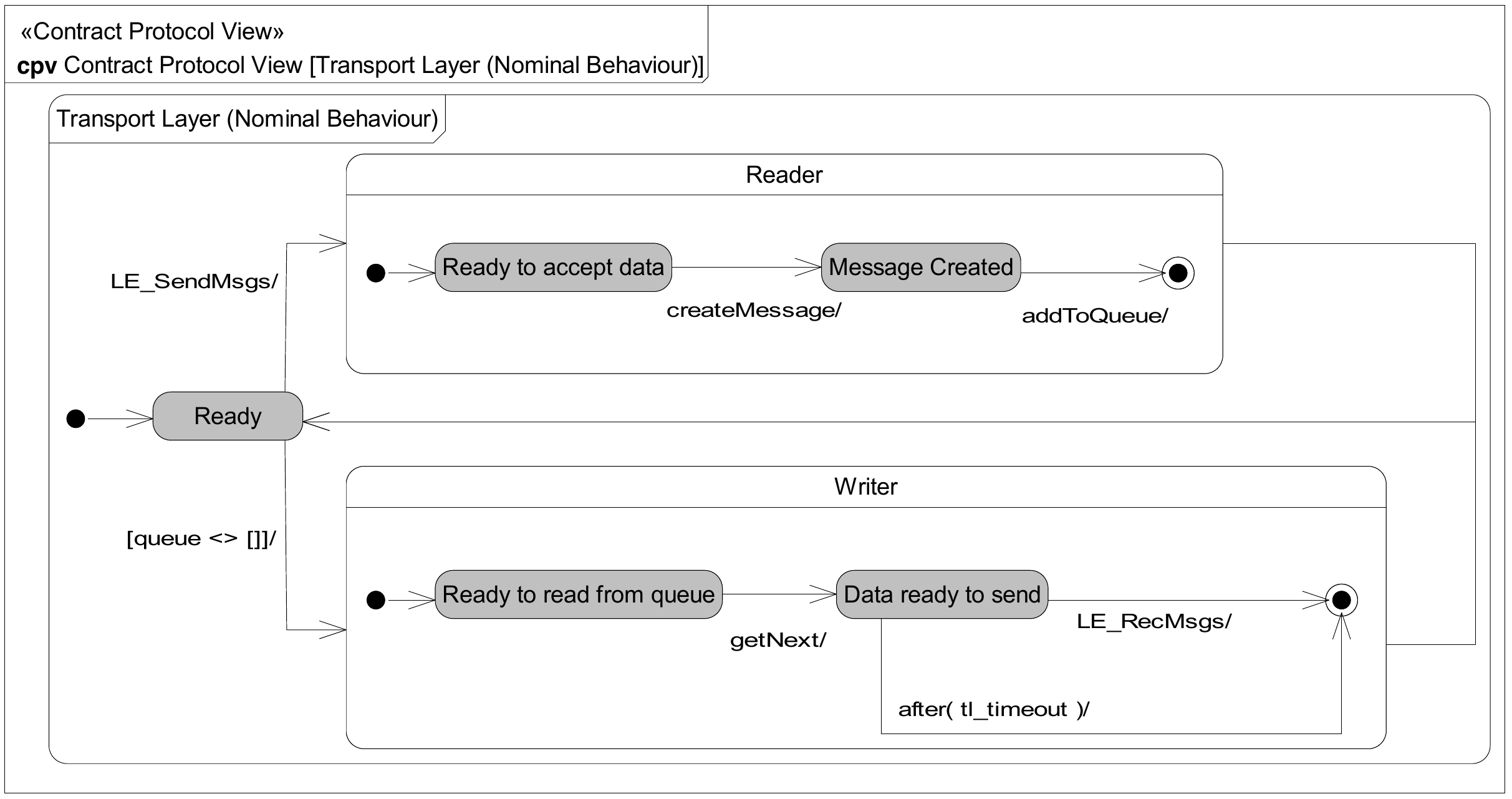}
	\caption{Contract Protocol View for the Transport Layer}
	\label{fig:cpv-nominal-tl}
\end{figure}

Figure~\ref{fig:cpv-nominal-tl} shows the state-machine of a well-behaved \emph{\ac{tl}} (we call the behaviour of a CS/SoS in the absence of faults its \emph{nominal} behaviour). The \emph{\ac{tl}} maintains a queue of messages in transit. When initialised, an \emph{LE Device} may send a message to the \emph{\ac{tl}} (\emph{LE\_SendMsgs}), in which case the \emph{\ac{tl}} enters the \emph{Reader} state. The received message
is packaged up  for the \emph{\ac{tl}} queue (by the operation \emph{createMessage}), added to the queue, and the \emph{Reader} state is exited. If the queue is not empty the \emph{\ac{tl}} can enter the \emph{Writer} state, in which a message is taken from the queue (operation \emph{getNext}) and the \emph{\ac{tl}} attempts to deliver it to its destination. The \emph{Reader} state is exited when either the delivery is successful, or it times out. 

\begin{figure}[h]
	\centering
	\includegraphics[width=0.49\textwidth]{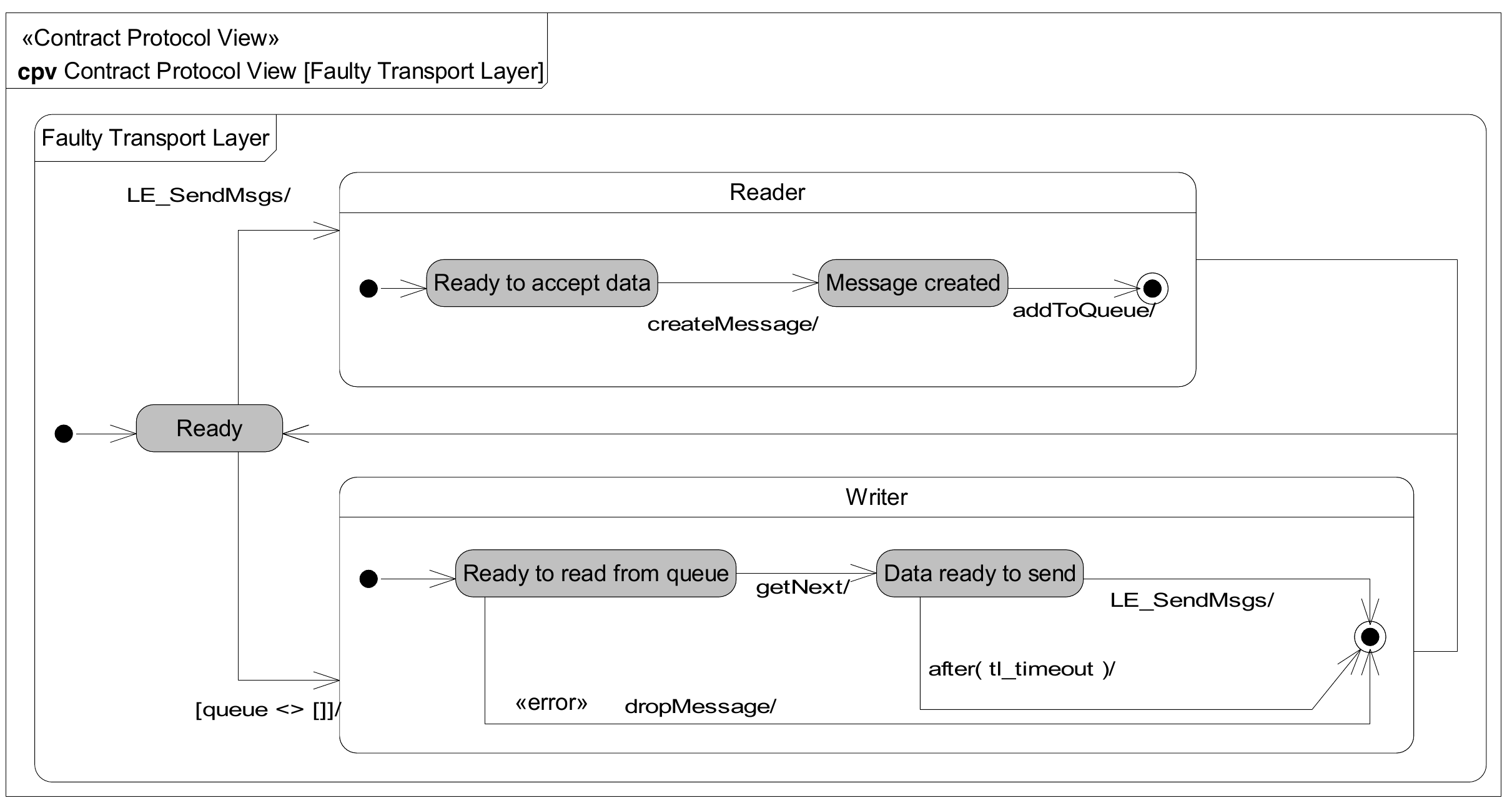}
	\caption{Contract Protocol View for the Faulty Transport Layer}
	\label{fig:cpv-faulty-tl}
\end{figure}

In this case study, we may consider the role of the SoS integrator, and thus observe that the \emph{\ac{tl}} in fact exhibits erroneous behaviour. The \emph{Faulty \ac{tl}} is depicted in Figure~\ref{fig:cpv-faulty-tl}. The \emph{Reader} state is identical, but the \emph{Writer} includes an erroneous transition \emph{dropMessage} (stereotyped as an \emph{$\langle\langle$error$\rangle\rangle$}) that  drops the message from the \emph{Ready to read from queue} state. This \emph{Faulty \ac{tl}} roughly corresponds to UDP in that no guarantees can be made over the delivery of messages.

In this paper we define the behaviour of a wrapper contract that can sit between an \emph{LE Device} and the \emph{\ac{tl}} in order to tolerate transient\footnote{A fault is \emph{transient} if its ``presence is bounded in time''~\cite{Avizienis2004}.  Note that we cannot tolerate \emph{permanent} \emph{TL} faults (whose presence are ``continuous in time''~\cite{Avizienis2004}) as these are indistinguishable from an \emph{LE Device} leaving the \ac{sos} (which is part of the nominal behaviour of the \ac{sos}).} faults in the \emph{\ac{tl}} as depicted in Figure~\ref{fig:cpv-faulty-tl}.  The use of a wrapper contract to ensure the local dependability of a pre-existing contract is a technique advocated by Romanovsky~\cite{Romanovsky2001}.

\begin{figure}[h]
	\centering
	\includegraphics[width=0.49\textwidth]{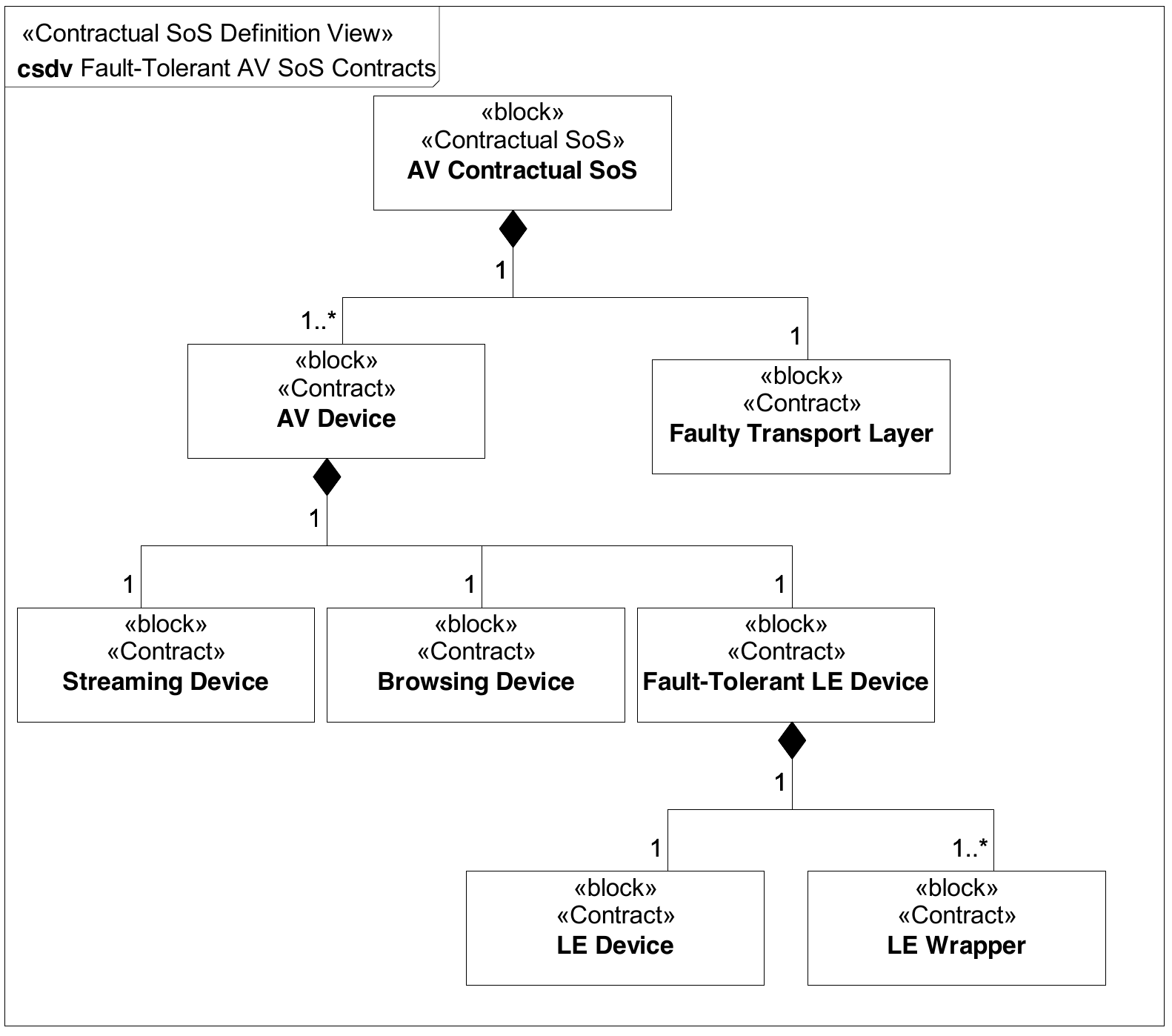}
	\caption{Contractual SoS Definition View of the Fault-Tolerant AV SoS}
	\label{fig:avsos}
\end{figure}

Figure~\ref{fig:avsos} shows the new \emph{contractual SoS} and the composition of the contracts in the SoS.

 Here, there is a single contract describing the behaviour of the \emph{Faulty Transport Layer}, and multiple copies of the \emph{AV Device} contract, which is in turn made up of three contracts, \emph{Streaming Device}, \emph{Browsing Device} and a \emph{Fault-Tolerant LE Device} contract adapted from the original \emph{LE Device} contract in Figure~\ref{fig:avsosnominal} to compensate for the faults exhibited by the \emph{\ac{tl}}.  The adaption is to compose the  basic \emph{LE Device} contract with a set of \emph{LE Wrappers} to create the \emph{Fault-Tolerant LE Device} contract. 

\begin{figure}[h]
	\centering
	\includegraphics[width=0.49\textwidth]{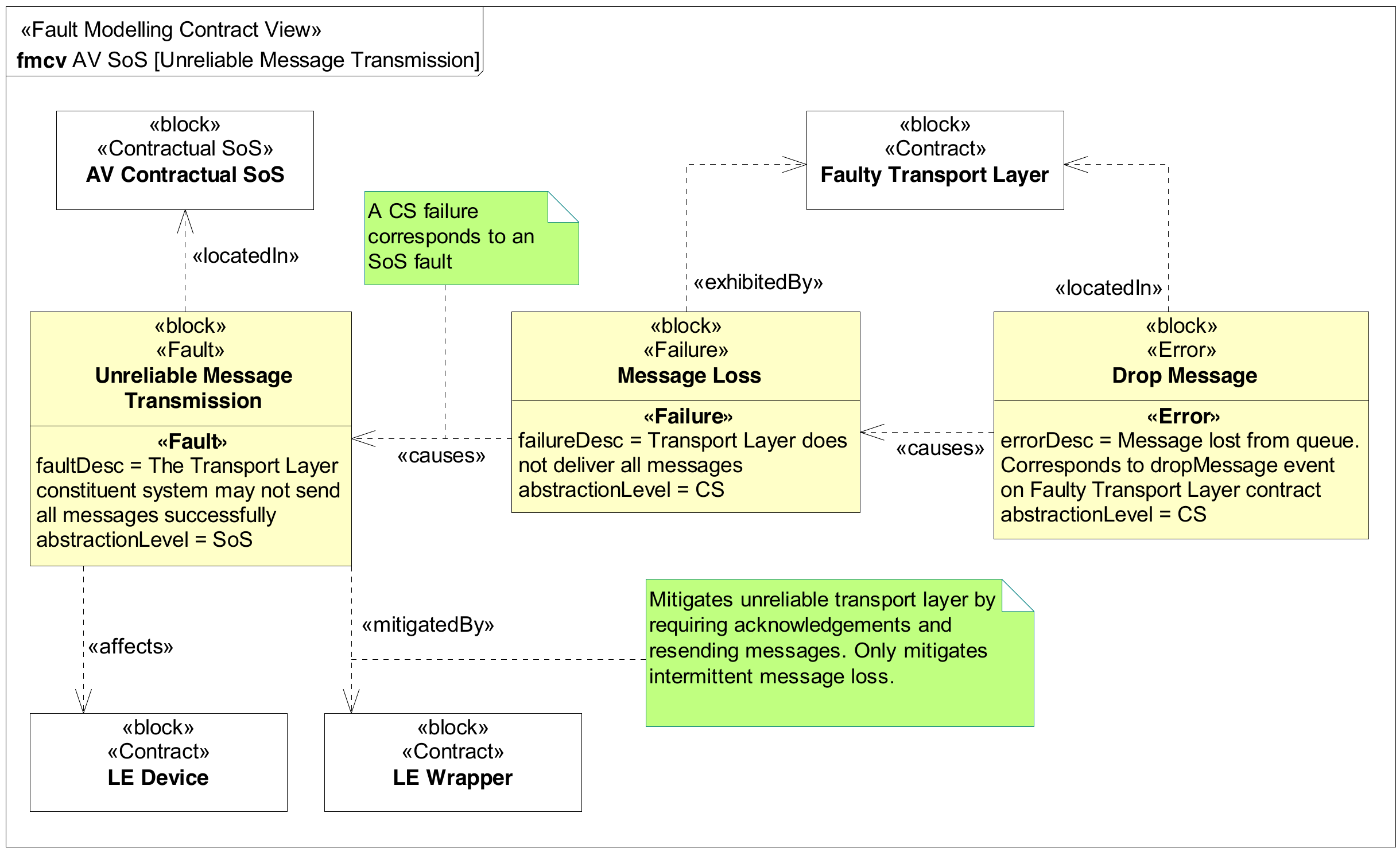}
	\caption{The Fault Modelling Contract View for the Unreliable Message Transmission fault}
	\label{fig:fmcv}
\end{figure}

The Fault Modelling Contract View in Figure~\ref{fig:fmcv} is an instantiation of a new viewpoint, extending the \contractspattern. The central three blocks are stereotyped as \emph{Fault}, \emph{Error} and \emph{Failure}, nomenclature standardised in~\cite{Avizienis2004} and used in the FMAF.  Recall, as described in Section~\ref{sec:fmaf}
that a failure of a CS can cause a fault of the SoS. 
 Each has a  textual description. The \emph{Drop Message} error and \emph{Message Loss} failures are artefacts at the \ac{cs}-level in the \emph{Faulty Transport Layer}. Based upon the dependability concepts used in the FMAF, the CS-level \emph{Message Loss} failure is considered to cause the \emph{Unreliable Message Transmission} fault at the SoS-level.  This fault, \emph{located in} the \emph{AV Contractual SoS}, \emph{affects} the \emph{LE Device} contract, and is \emph{mitigated by} the \emph{LE Wrapper} contract.

\begin{figure}[h]
	\centering
	\includegraphics[width=0.49\textwidth]{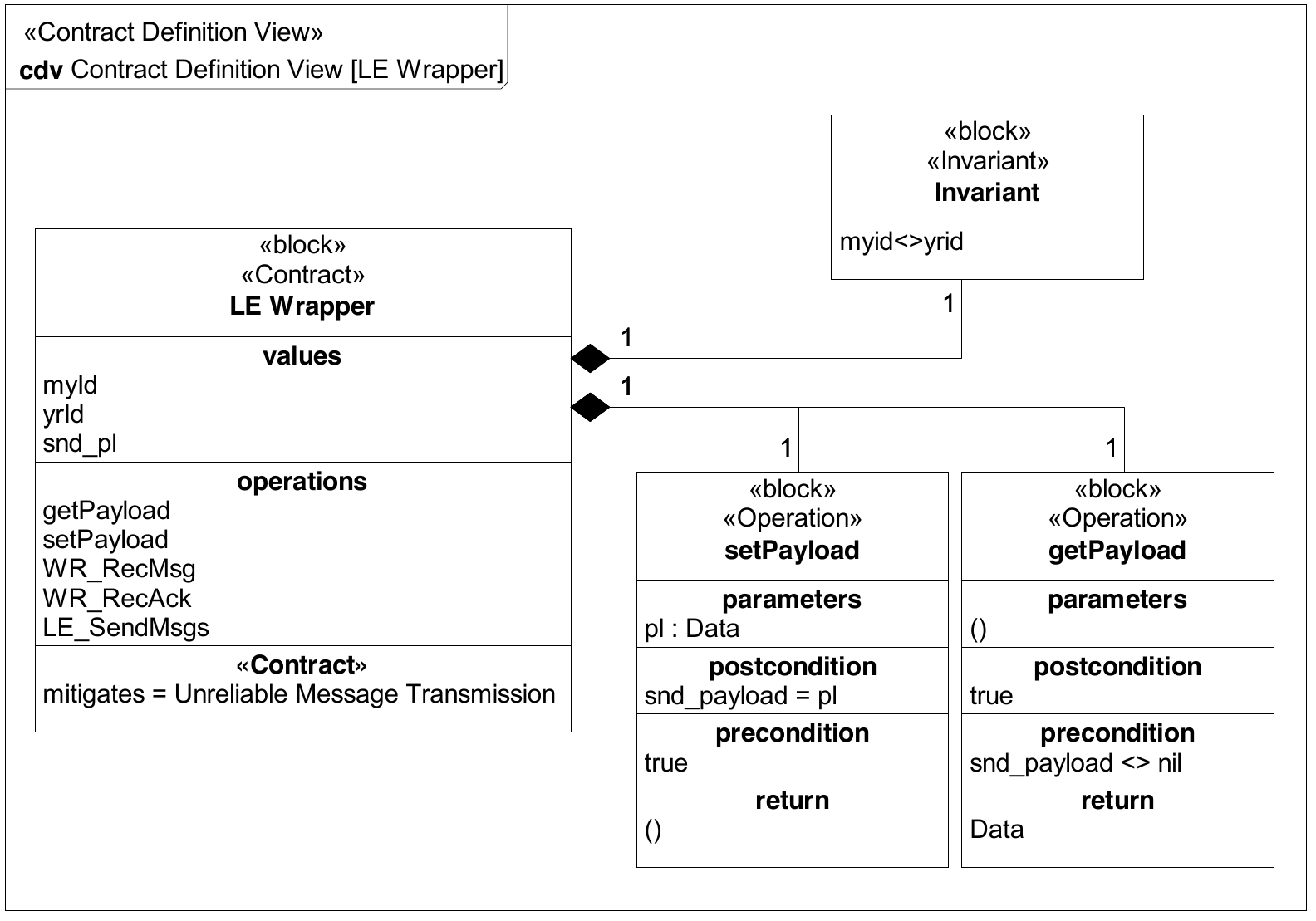}
	\caption{The Contract Definition View of the LE Wrapper}
	\label{fig:cdv-wrapper}
\end{figure}

The purpose of the \emph{LE Wrapper} contract is to acknowledge receipt of messages destined for the \emph{LE Device} contract and to listen for acknowledgments to messages that it has sent.  The Contract Definition View in Figure~\ref{fig:cdv-wrapper} gives the state,  operations and invariants of the \emph{LE Wrapper}. Each  \emph{Fault-Tolerant LE Device} consists of an
\emph{LE Device} and a wrapper for every other \emph{LE Device} with which it communicates.  The contract for \emph{LE Wrapper} is therefore parameterised by \emph{myId} (the identity of the ``owning'' \emph{LE Device}) and \emph{yrId} (the identity of the receiving \emph{LE Device}.) The \emph{snd\_pl} value holds the payload that is in the process of being sent at a given time. The figure shows an extension made to the \contractspattern\ -- the inclusion of a \emph{mitigates} tag, which explicitly records that the \emph{LE Wrapper} contract mitigates the \emph{Unreliable Message Transmission} SoS-level fault, as identified in Figure~\ref{fig:fmcv}.

\begin{figure}[h]
	\centering
	\includegraphics[width=0.49\textwidth]{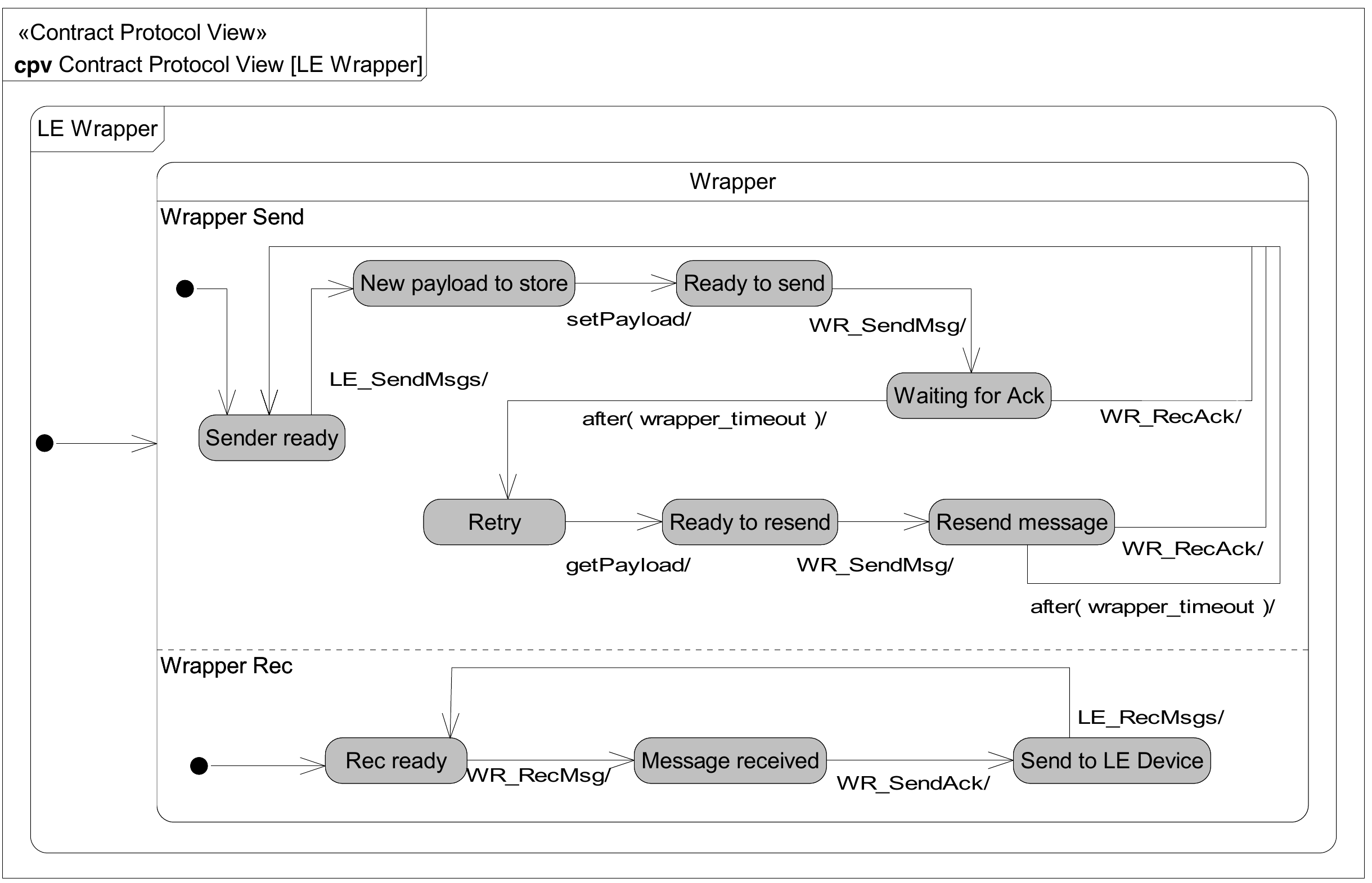}
	\caption{The Contract Protocol View for the LE Wrapper}
	\label{fig:cpv-wrapper}
\end{figure}

 The required behaviour of the \emph{LE Wrapper} contract is specified in Figure~\ref{fig:cpv-wrapper}. \emph{Wrapper Send} and \emph{Wrapper Rec} are parallel states.  \emph{Wrapper Send} passes on messages from the owning \emph{LE Device} to the \emph{\ac{tl}}.
It begins by storing the message, in case it needs to be resent, then sends it. If an acknowledgment is received (within the time limit \emph{wrapper\_timeout}) the wrapper goes back to its initial state, otherwise it tries once again to resend the stored message. 
This resend only happens once; even if an acknowledgment is not received after the second attempt, the wrapper does not attempt to resend\footnote{The number of retries can be trivially increased, but the AV SoS needs to be tolerant to devices exiting the SoS at any time, therefore the \emph{LE Wrapper} should not retry to send the message indefinitely.  The exact choice on the number of retries is a trade-off between maintaining consistency within the SoS in the presence of faults and reducing unnecessary traffic on the \emph{TL}.}. 
\emph{Wrapper Rec} receives and acknowledges incoming messages from the \emph{\ac{tl}} and passes them on to the \emph{LE Device}\footnote{This \emph{LE Wrapper} is therefore similar to, but considerably simpler than, protocols for making UDP reliable such as Automatic Repeat Request (ARQ) and  Pragmatic General Multicast (PGM).}. 

In summary, Table~\ref{tab:pattern_changes} identifies those viewpoints of the \contractspattern\ which have been extended to include fault modelling concepts (extensions are shown in italics) and the additional Fault Modelling Contract Viewpoint presented in this paper.

\begin{table}
\caption{Informal description of the \fmconpatt\ viewpoints}
\label{tab:pattern_changes}
\begin{center}
\begin{tabular}{|l|l|}	
  \hline
   \multicolumn{2}{|c|}{\textbf{Existing Viewpoints}}\\
	\hline
	\hline
  \textbf{Name} & \textbf{Purpose of Viewpoint}\\
    \hline
      Contractual SoS &Identifies the contracts which comprise the\\
      Definition Viewpoint  &  Contractual SoS.\\
    \hline
      Contract Conformance & Denotes the contracts to which the SoS constituent \\
      Viewpoint & systems conform.\\ 
    \hline
      Contract Connections &Shows connections and interfaces between contracts \\
      Viewpoint &  of the Contractual SoS. \\
    \hline
      Contract Definition & Defines the operations, state variables and\\
      Viewpoint &  state invariants for a single contract. \\
                &\emph{Failure modes and mitigated faults added.}\\
	\hline
	  Contract Protocol & Protocol specification of a contract.\\
	  Viewpoint & \emph{Erroneous transitions added and fault-tolerant}\\
	            & \emph{protocols supported.}\\
	\hline
	\hline
   \multicolumn{2}{|c|}{\textbf{New Viewpoint}}\\
	\hline
	\hline
  \textbf{Name} & \textbf{Purpose of Viewpoint}\\
    \hline
    	  Fault Modelling & Associates faults, errors and failures with\\
	  Contract Viewpoint & contracts. Includes \emph{exhibited by}, \emph{mitigated}\\ 
	                     & \emph{by} and \emph{affected by} relationships.\\
	\hline
\end{tabular}
\end{center}
\end{table}

\section{Relationship with FMAF}
\label{sec:fm-contracts-fmaf}


As described in Section~\ref{sec:background}, the \contractspattern\ and FMAF are defined using a common framework. We may therefore make extensions (as is done with the \fmconpatt \ in this paper) and also combine them using their ontologies to identify shared modelling elements and relationships. In this section, we briefly consider the areas in which the \contractspattern\ and FMAF overlap and consider one possible area in which there is a difference in the proposed modelling approach.

The \fmconpatt\ and the FMAF can be used in combination to provide a consistent and complementary set of viewpoints of an \ac{sos} model.  Both aspects of the model should have a consistent understanding of the faults, errors and failures of the \ac{sos}.  To achieve this we recommend that a single definition of these is used for both sets of viewpoints, by making use of the FMAF Fault/Error/Failure Definition Viewpoint (FEFDV) \cite{Andrews2014} within the \fmconpatt.  The FEFDV provides details of all of the relevant faults, errors and failures of the SoS and identifies any relationships that exist between them. The Fault Modelling Contract View in Figure~\ref{fig:fmcv}, must therefore be consistent with the FEFDV.

The structural aspects of the patterns are complementary -- where the FMAF supports definition of the composition and connections of \acp{cs}, the \fmconpatt \ does the same for contracts.  The Contract Conformance Viewpoint of the \fmconpatt \ relates the \acp{cs} to the contracts and can thus be used as a basis for checking the consistency of these complementary aspects of the model.

The approaches to modelling behaviour in the FMAF and in the \fmconpatt \ differ quite considerably.  Whilst the FMAF uses sequence diagrams and activity diagrams to focus on processes of the \ac{sos} and interactions between \acp{cs}, the \fmconpatt \ uses state machines to target the states and transitions of single \acp{cs} (via the definition of the contract protocols they are required to adhere to).  These two complementary approaches can be effectively used side by side to provide a more complete treatment of the behaviour of the \ac{sos}.

Finally, traceability links could be used as a semi-formal way (see for example \cite{Andrews2014}) of relating the complementary aspects of the \fmconpatt \ and the FMAF -- the implementation of this is the subject of further research.



\section{Conclusions}
\label{sec:conclusions}

%

%
%

In this paper we have highlighted the crucial need to examine the impact of faults when defining contracts for \acp{cs}, in order to create a fault-tolerant  \ac{sos}.  
Using an Audio Visual \ac{sos} case study we have illustrated how the previously developed \contractspattern~\cite{Bryans2014} can be extended to include consideration for faults and fault tolerance. 
Extending the \contractspattern\ in this way has been shown to be complementary to our previous efforts in defining an architectural framework (the FMAF \cite{Andrews2014}) for incorporating faults at the architectural design phase of creating an \ac{sos}.

Fault modelling has been an active area of research for some time and there is a rich taxonomy of fault types, as well as a rich set of ``countermeasures'' (fault tolerance and recovery, mitigation, etc)~\cite{Avizienis2004}.
In this paper we have used a single fault and recovery mechanism in order to investigate the feasibility of integration of the \contractspattern\ and the FMAF. 
A natural continuation of this work is to extend the types of faults and countermeasures that we cover, as well as considering the interaction of multiple faults. 
Our preliminary results may also be strengthened by their application to a wider range of \acp{sos}, leading to a more complete set of extensions to the \contractspattern \ for the purpose of modelling faults and fault tolerance within \acp{sos}.  
The integration of the \fmconpatt \ with the FMAF also deserves a more thorough treatment, and we consider two main directions: the first of which relates to the first piece of future work mentioned earlier -- identifying new viewpoints that may be needed to support a wider variety of faults and countermeasures, along with new relationships between such viewpoints and the FMAF; and the second is the recording of traceability links (as discussed in Section~\ref{sec:fm-contracts-fmaf}) between elements of the \fmconpatt \ and elements of the FMAF (building upon the work in~\cite{Andrews2014}).

Further types of extension to the \contractspattern\ could be envisaged.  %
Ongoing work on static fault analysis of an \ac{sos} investigates  marking up structural architectural models with certain fault information.  Analysis can then be carried out using an industrial fault analysis tool.  It seems natural to incorporate the mark-up language within the \fmconpatt.

\section*{Acknowledgements}

The work presented here is supported by the EU Framework 7 Integrated Project ``Comprehensive Modelling for Advanced Systems of Systems'' (COMPASS, Grant Agreement 287829) (for more information see \url{http://www.compass-research.eu}) and the EPSRC Platform Grant on Trustworthy Ambient Systems (TrAmS-2).


\bibliographystyle{IEEETran}
\bibliography{edsos_le_fm}

\end{document}